\documentclass[conference]{IEEEtran}
\IEEEoverridecommandlockouts
\usepackage{}
\usepackage{xcolor}
\usepackage{color}
\usepackage{amsmath}
\usepackage{graphicx}
\usepackage{amssymb}
\usepackage{amsmath,amsthm} 
\usepackage{float}
\usepackage{enumerate}
\usepackage{epstopdf}
\usepackage{dblfloatfix}
\usepackage{caption}
\usepackage{textgreek}
\usepackage{amsfonts}
\usepackage{fancyhdr}
\usepackage{bbm}
\usepackage{cases}
\usepackage{cite}
\usepackage{breqn}
\usepackage{multirow}
\usepackage{mathrsfs}
\usepackage{algpascal}
\usepackage{algc}
\usepackage{algorithmicx}
\usepackage[ruled]{algorithm}
\usepackage{algpseudocode}
\usepackage{graphics}
\usepackage{epsfig}
\usepackage{mathrsfs} 
\usepackage{dsfont}

\usepackage{appendix}

\floatname{algorithm}{Algorithm}

\algdef{SE}[DOWHILE]{Do}{doWhile}{\algorithmicdo}[1]{\algorithmicwhile\ #1}%

\ifCLASSINFOpdf
 \else
\fi

\begin{document}
\theoremstyle{definition} 
\newtheorem{theorem}{Theorem}[section]
\newtheorem{definition}[theorem]{Definition}
\newtheorem{lemma}[theorem]{Lemma}
\newtheorem{example}[theorem]{Example}
\newtheorem{Proposition}[theorem]{Proposition}
\newtheorem{Corollary}[theorem]{Corollary}

\title{Flexible Intelligent Metasurface-Aided ISAC: User Fairness Optimization and Performance Evaluation}

\author{
 \IEEEauthorblockN{Hailun Huang\IEEEauthorrefmark{2}, 
        Yuwen Cao\IEEEauthorrefmark{2}, 
         Jiguang He\IEEEauthorrefmark{3},
         Tomoaki Ohtsuki\IEEEauthorrefmark{4}
    }
    \IEEEauthorblockA{\IEEEauthorrefmark{2}School of Information and
Intelligent Science, Donghua University, Shanghai, China}
\IEEEauthorblockA{\IEEEauthorrefmark{3}School of Computing and Information Technology, Great Bay University, Dongguan, China}
\IEEEauthorblockA{\IEEEauthorrefmark{4}Department of Information and Computer Science, Keio University, Yokohama, Japan}
}

\markboth{IEEE WIRELESS COMMUNICATIONS LETTERS,~Vol.~XX, No.~XX, XXX~2019}
{}

\maketitle

\begin{abstract}
This paper investigates max-min user fairness optimization for flexible intelligent metasurface (FIM) and non-orthogonal multiple access (NOMA)-assisted integrated sensing and communication (ISAC) systems with active self-localization. 
To tackle the multi-user interference, performance imbalance, and neglected sensing accuracy problems encountered in conventional ISAC designs, 
we derive the closed-form Cramér-Rao lower bound (CRLB) for angle-of-departure (AoD) estimation in target sensing and embed it into a max-min fairness optimization framework. The optimization problem, which jointly designs the base station transmit beamforming, FIM reflection coefficients, and surface deformation, is non-convex and solved by an alternating optimization (AO) algorithm. Notably, the devised optimization framework facilitates superior performance
trade-off in terms of spectrum resource utilization
between communication and sensing tasks.
Simulation results validate that the proposed scheme significantly improves user fairness, balances communication performance and sensing precision effectively, and reveals the coupling characteristic between signal-to-interference-plus-noise ratio (SINR) and sensing CRLB. This work provides a feasible solution for FIM-aided ISAC system optimization in 6G networks.
\end{abstract}


\section{Introduction}
The 6G wireless networks are envisioned to support the seamless coexistence of ultra-fast multi-user communication and high-precision environmental sensing, giving rise to ISAC as a core enabling technology\cite{2016Next}, \cite{11575614}. Driven by the demand for smart and reconfigurable radio propagation environments, reconfigurable intelligent surfaces (RIS) have emerged as a disruptive low-cost and energy-efficient solution for ISAC systems\cite{11145277}. Conventional RIS architectures rely on rigid, planar passive elements, which limits their adaptability to non-line-of-sight (NLoS) obstructions. 
In contrast, flexible intelligent metasurfaces (FIMs), as an advanced RIS paradigm, feature mechanically deformable elements with independent displacement along the normal direction \cite{2020Smart}.

Despite the promising potential of FIM-aided ISAC systems, several critical challenges remain unresolved in the open literature. Firstly, most existing RIS-ISAC designs focus on rigid surfaces and overlook the unique degrees of freedom (DoFs) brought by FIM deformation \cite{2021Cram}, 
leading to suboptimal performance in practical deployment. Secondly, multi-user interference becomes severe in dense communication user (CU) scenarios, causing significant performance imbalance and violating user fairness constraints. Although NOMA with successive interference cancellation (SIC) has been widely adopted to mitigate intra-system interference\cite{2023Digital}, the joint design of NOMA-SIC and FIM-aided ISAC under max-min fairness criteria remains largely unexplored. Thirdly, sensing performance characterized by the estimation accuracy of angle-of-arrival (AoA) and propagation distance is often treated as a secondary objective or ignored in fairness-oriented optimization. The Fisher information matrix and CRLB provide a fundamental limit for unbiased parameter estimation\cite{2024Cram}, yet few studies integrate CRLB-based sensing accuracy constraints into the fairness optimization framework. Furthermore, practical hardware limitations including FIM deformation bounds, total transmit power budgets, and SIC decoding order constraints are frequently oversimplified, resulting in physically infeasible optimization solutions.

Against the above challenges, this paper constructs an FIM-aided NOMA-ISAC system with active self-localization realized by transmitting positioning reference signals (PRS), and optimizes BS beamforming, FIM reflection coefficients and surface deformation jointly under max-min fairness. Notably, the main contributions of this work are summarized as follows:

\begin{itemize}
\item We construct an FIM-assisted ISAC system model with active positioning via transmitting communication signals and PRS. CUs receive reflected signals for NOMA communication, while the sensing user (SU) receives PRS for active positioning.
\item We derive the closed-form expressions of the Fisher information matrix and corresponding CRLB for the key sensing parameters, i.e., the two-dimensional AoD from the SU $\theta_{ft}$  and $\varphi_{ft}$ 
which quantify the theoretical lower bound of unbiased estimation accuracy. The derived CRLB is further adopted as a critical constraint to guarantee the sensing performance in the fairness optimization.
\item We validate the effectiveness of the proposed joint optimization framework through theoretical analysis, demonstrating that the integration of FIM deformation, 
and CRLB constraints significantly improves user fairness while satisfying strict communication and sensing performance requirements.
\end{itemize}

\section{System Model}
\begin{figure}
    \centering
 \setlength{\abovecaptionskip}{-0.6cm}
\includegraphics[width=0.5\textwidth]{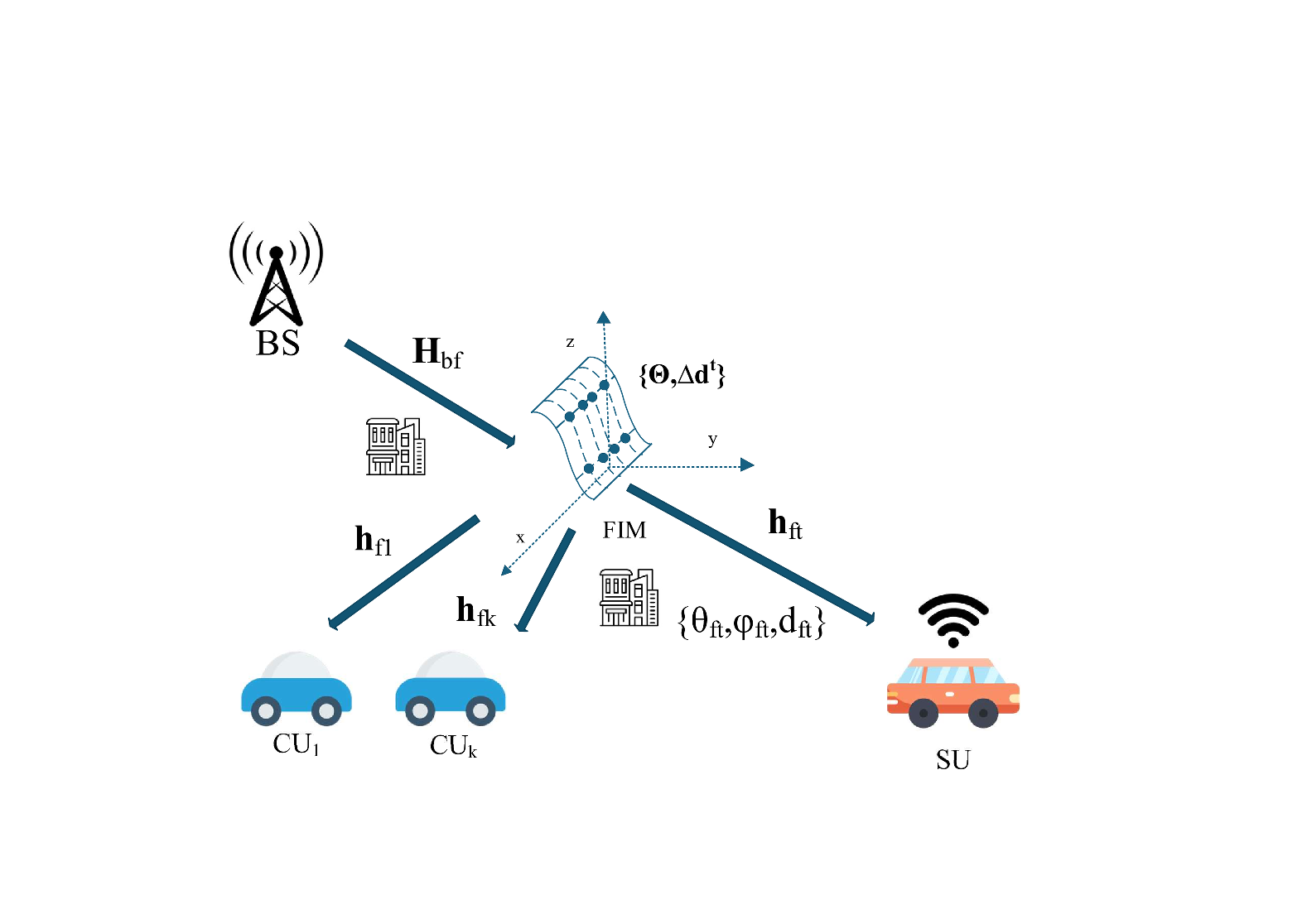}
    \caption{An illustration of the proposed FIM-assisted ISAC system.}
    \label{fig:placeholder01}
\end{figure}

%
Consider an FIM-assisted ISAC system consisting of a BS, a pure reflection mode FIM, $K$ CUs, and one SU. Fig. \ref{fig:placeholder01} illustrates the schematic of the proposed FIM assisted downlink ISAC system. 
The BS with a $M$-antenna uniform linear array (ULA) simultaneously transmits
data to $K$ single-antenna CUs and sends
positioning reference signals for accurate AoD and distance estimation at the SU. 
The FIM contains $N = N_x \times N_z$ passive elements configured as a uniform planar array (UPA) on the $x$–$z$ plane. Unlike conventional rigid RIS, FIM supports independent mechanical deformation along the normal $y$-axis and operates in a single mode to realize simultaneous signal reflection. The $K$ CUs receive communication signals via the reflection link of the FIM and employ SIC to suppress interference from the sensing signal. Meanwhile, SU obtains the sensing signal via the transmission link from FIM to SU.

\subsection{FIM Model}
The position of the $n$-th FIM unit is denoted as $\mathbf{p}_{n} = \left[ x_{n},\, y_{n},\, z_{n} \right]^{\mathrm{T}} \in \mathbb{R}^{3}$. Herein, $x$ and $z$ coordinates are fixed by letting $x_{n} = d_{x} \times \bmod(n-1, N_{x})$ and $z_{n} = d_{z} \times  [(n-1)/N_{x}] $, where $d_x=λ/2$ and $d_z=λ/2$ are the element spacings, and $λ$ is the carrier wavelength. The deformation on the $y$-axis satisfies $\left| \Delta d_{n}^{t} \right| \leq d_{\max}^{t},\ \forall n$, and $\Delta d_{n}^{t}$ denotes the deformation displacement of the $n$-th element of the FIM. 
The diagonal reflection coefficient matrix are $\boldsymbol{\Theta}_{}=\operatorname{diag}\left(e^{j\phi_{1}},\dots,e^{j\phi_{N}}\right)$, and the phase shifts satisfy $\phi_{n}\, \in (0, 2\pi],\quad \forall n\in\{1,\ldots,N\}$.

\subsection{Channel Model}
We assume that the FIM-empowered ISAC system operates at mmWave frequency band. 
Hence, we adopt the Saleh-Valenzuela parametric channel model to construct the involved
three individual channels, as shown in Fig. \ref{fig:placeholder01}. 
Assume that $K$ CUs are located at $\{(\theta_k,\varphi_k)\}_{k=1}^K$, with $\theta_k\in(0,\pi]$ and $\varphi_k \in (0,\pi]$ denoting the elevation and azimuth angles of CU $k$ with respect to FIM, respectively. Let 
$\theta_{bf}$ and $\varphi_{bf}$ be the  elevation and azimuth AoD of the target with respect to the
BS. Then, the 
transmitting steering vector $\boldsymbol{\alpha}_1(\theta_{bf}) \in\mathbb{C}^{M\times 1}$ with respect to the BS can be expressed as
\begin{equation} \small
\boldsymbol{\alpha}_1(\theta_{bf}) = \biggl[
e^{-j \frac{2\pi d_x}{\lambda} \frac{M-1}{2} \sin\theta_{bf}},\,
\ldots,\,
e^{j \frac{2\pi d_x}{\lambda} \frac{M-1}{2} \sin\theta_{bf} }
\biggr]^{\mathrm{T}},
\end{equation}
where $d_x$ denotes the inter-element spacing, which is set to be lossless at half wavelength without loss of generality. $\lambda$ represents the carrier’s wavelength.
In addition, the 
transmitting steering vector $\boldsymbol{\alpha}_i(\theta_k, \varphi_k, \mathbf{\Delta d}^{t}) \in \mathbb{C}^{N\times 1}$, $i=\{2,3,4\}$, with respect to the FIM can be expressed as\cite{11060929}
\begin{equation} \small
\boldsymbol{\alpha}_i(\theta_k, \varphi_k, \mathbf{\Delta d}^{t}) = [ 
 \boldsymbol{\alpha}_{z}(\theta_k)\otimes \boldsymbol{\alpha}_x(\theta_k,\varphi_k) ] \odot \boldsymbol{\alpha}_y(\theta_k,\varphi_k,\mathbf{\Delta d}^{t}).
\end{equation}
Steering vectors $\mathbf{a}_{z}(\theta_{k})$, $\mathbf{a}_{x}(\theta_{k},\varphi_{k})$, and $\boldsymbol{\alpha}_y(\theta_k,\varphi_k,\mathbf{\Delta d}^{t})$ are respectively defined as\cite{11060929}
\begin{align}
    \label{eq:002}
    \mathbf{a}_{z}(\theta_{k})=
\begin{bmatrix}
e^{-j\frac{2\pi d_z}{\lambda}\frac{N_z -1}{2}\cos\theta_k},\ldots,e^{j\frac{2\pi d_z}{\lambda}\frac{N_z-1}{2}\cos\theta_k}
\end{bmatrix}^{T},
\end{align}

\begin{align}
    \label{eq:003}
\begin{split}
    \mathbf{a}_{x}(\theta_{k},\varphi_{k})=
\Big[&
e^{-j\frac{2\pi d_x}{\lambda}\frac{N_x -1}{2}\sin\theta_k\cos\varphi_k},\\&\ldots,e^{j\frac{2\pi d_x}{\lambda}\frac{N_x -1}{2}\sin\theta_k\cos\varphi_k}
\Big]^{T},
\end{split}
\end{align}


\begin{align}
\label{eq:004}
\begin{split}
\mathbf{a}_{y}(\theta_{k},\varphi_{k},\Delta\mathbf{d}^{t})  =\Big[&e^{-j\frac{2\pi}{\lambda}\Delta d_{1}^{t}\sin\theta_{k}\sin\varphi_{k}}, \\
 & \ldots,e^{-j\frac{2\pi}{\lambda}\Delta d_{N_{}}^{t}\sin\theta_{k}\sin\varphi_{k}}\Big]^{T}.
\end{split}
\end{align}

Then, the channel vector $\mathbf{h}_{fk}(\mathbf{\Delta d}^{t})$ between the FIM and the CU $k$ can be expressed as 
\begin{equation} \small
\mathbf{h}_{fk}(\mathbf{\Delta d}^{t})   =\frac{e^{-\frac{j2\pi d_{fk}}{\lambda}}}{\sqrt{\rho_{fk}}} \boldsymbol{\alpha}_2(\theta_{k}, \varphi_{k}, \mathbf{\Delta d}^{t}) ,
\end{equation}
where $d_{fk}$ and $\rho_{fk}$ represent the distance and path loss between the
FIM and the CU $k$, $k\in\{1,\ldots,K\}$, respectively.
For simplicity, in our
system we consider free-space path loss \cite{2024Cram}, which is modeled by 
letting 
$
\rho_{fk} = d_{fk}^2f_c^2/10^{8.755}, 
$
with $f_c$ being the carrier frequency.
Likewise, the channel vector $\mathbf{h}_{ft}(\mathbf{\Delta d}^{t})$ between the FIM and the SU can be represented as 
\begin{equation} \small
\mathbf{h}_{ft}(\mathbf{\Delta d}^{t})  =\frac{e^{-\frac{j2\pi d_{ft}}{\lambda}}}{\sqrt{\rho_{ft}}}\cdot \boldsymbol{\alpha}_3(\theta_{ft},\varphi_{ft},\mathbf{\Delta d}^{t}),
\end{equation}
where $d_{ft}$ and $\rho_{ft}$ denote the distance and path loss between the FIM and the SU, respectively. Besides,  $\rho_{ft}$  follows the same  assumption as that of $\rho_{fk}$. Besides, $\theta_{ft}$ and $\varphi_{ft}$ correspond to the the elevation and azimuth AoD of SU with respect to the FIM, respectively. 
Similar to $\mathbf{h}_{ft}(\mathbf{\Delta d}^{t})$,
the wireless channel matrix $\mathbf{H}_{bf}(\mathbf{\Delta d}^{t})\in\mathbb{C}^{N\times M}$ between BS and FIM can be expressed as follows:
\begin{equation} \small
\mathbf{H}_{bf}(\mathbf{\Delta d}^{t})    =\frac{e^{-\frac{j2\pi d_{bf}}{\lambda}}}{\sqrt{\rho_{bf}}}\cdot \boldsymbol{\alpha}_4(\theta_{bf},\varphi_{bf},\mathbf{\Delta d}^{t}) \boldsymbol{\alpha}^H_1(\theta_{bf}),
\end{equation}
where $d_{bf}$ corresponds to the distance between the BS and the FIM. $\rho_{bf}$ follows the same assumption as that of $\rho_{fk}$ and $\rho_{ft}$.

\subsection{Signal Model}
We consider a slotted transmission structure with a total of $L$ time slots, indexed by $l=1,\ldots,L$. The BS transmits a dedicated PRS (i.e., pilot) in each time slot for active self-localization, and optimizes the sensing beamformer $\mathbf{w}_{t,l}$, FIM reflection coefficient matrix $\boldsymbol{\Theta}$, and surface deformation $\mathbf{\Delta d}^t$ independently for each slot $l$, i.e.,
\begin{equation}
\small
    \mathbf{x}_l = \sum\nolimits_{k=1}^{K} \mathbf{w}_{k} s_{k,l} + \mathbf{w}_{t,l} \varphi_l\label{9},
\end{equation}
where $\mathbf{w}_{k} \in \mathbb{C}^{M \times 1}$ denotes the communication beamforming vector of the $k$-th CU, and $s_{k,l}$ represents the corresponding communication data symbol at the $l$-th time slot. $\mathbf{w}_{t,l} \in \mathbb{C}^{M \times 1}$ is the slot-wise optimized sensing beamforming vector for the $l$-th slot, and $\varphi_l$ represents the normalized orthogonal PRS pilot symbol at slot $l$, to ensure that the signals over $L$ time slots are free from mutual interference and enable CRLB calculation via superposition. Besides, $\varphi_l$ satisfies two standard mathematical constraints: i) normalization and ii) inter-slot orthogonality, satisfying $\mathbb{E}\left[|\varphi_l|^2\right] = 1$ and $\mathbb{E}\left[\varphi_l \varphi_l^*\right] = 0, \forall k \neq l$, with $\varphi_l^*$ denoting the complex conjugate of the positioning pilot $\varphi_l$ in the $l$-th time slot.

The received signal at the $k$-th CU in the $l$-th time slot consists of the desired communication signal, interference from other CUs, interference induced by the SU sensing signal, as well as the Gaussian noise, which can be represented as
\begin{equation}
\small
\begin{split}
    y_{k,l} = \mathbf{h}_k \mathbf{w}_{k} s_{k,l} + \sum\nolimits_{\substack{m=1, m \neq k}}^{K} \mathbf{h}_k \mathbf{w}_{m} s_{m,l} + \mathbf{h}_k \mathbf{w}_{t,l} \varphi_l + n_k,
    \end{split}
\end{equation}
where $\mathbf{h}_k = \mathbf{h}_{fk}^{H}(\mathbf{\Delta d}^{t}) \boldsymbol{\Theta}\mathbf{H}_{bf}(\mathbf{\Delta d}^{t}) \in\mathbb{C}^{1\times M}$ is the cascaded channel of the BS-FIM-CU $k$ link, and
$n_k \sim \mathcal{CN}(0, \sigma_k^2)$ is additive white Gaussian noise (AWGN). Subsequently, the sensing signal interference is eliminated and the desired signal for CUs is recovered consecutively by using SIC technique. The received signal at the SU in the $l$-th time slot contains the combined communication and sensing signals, as well as Gaussian noise, which is formulated as:
\begin{equation}
    y_{t,l} = \mathbf{g}_s \left( \sum\nolimits_{k=1}^{K} \mathbf{w}_{k} s_{k,l} + \mathbf{w}_{t,l} \varphi_l \right) + n_t,
\end{equation}
where $\mathbf{g}_s = \mathbf{h}_{ft}^{H}(\mathbf{\Delta d}^{t}) \boldsymbol{\Theta}\mathbf{H}_{bf}(\mathbf{\Delta d}^{t}) \in\mathbb{C}^{1\times M}$ is the cascaded
channel of the BS-FIM-SU link, and
$n_t \sim \mathcal{CN}(0, \sigma_t^2)$ is the AWGN at the SU. Based on the measures of the received PRSs, i.e., $\mathbf{y}_t = [y_{t,1}, y_{t,2}, \dots, y_{t,L}]^T \in \mathbb{C}^{L \times 1}$, the SU estimates the channel parameters ($\theta_{ft},\varphi_{ft}$) for active self-localization task. Based on the stacked PRS observations $\mathbf{y}_t$, the SU jointly estimates the key sensing parameters $\theta_{ft}$ and $\varphi_{ft}$ to realize accurate active self-localization. \footnote{
Deep learning-guided low-overhead beam and power allocation approaches based on received signal measurements are presented in \cite{10146432} and \cite{9314253}. 
Extending the established optimization framework with tailored deep learning techniques for accurate active self-localization will be explored in our future work.
}

\section{Performance Metrics}
%

\subsection{Communication Performance Metric}
We adopt the SINR as the
communication performance metric.
According to the NOMA principle \cite{2016Power}, \cite{2017A}, users with favorable channel conditions employ SIC to decode the messages of users with weaker signals before recovering their own information. In our FIM-aided NOMA-ISAC system, the BS concurrently transmits communication data for CUs and PRS for the SU. For all CUs, the sensing signal acts as undesired interference that will degrade communication quality and multi-user fairness. To fully eliminate sensing-induced interference before recovering communication signals, the sensing signal must be decoded first and then removed via SIC at each CU. Following the core rule of power-domain NOMA, a signal with higher received power will be decoded earlier. For this reason, we purposely set the received power of the sensing signal to be higher than that of all communication signals. Without loss of generality, the channel gains of all users are assumed to follow the order as 
$0 \leq \left\| \mathbf{h}_1 \right\|^2 \leq \cdots \leq \left\| \mathbf{h}_k \right\|^2 \leq \cdots\left\| \mathbf{h}_K \right\|^2$ \cite{2024Cram},
where $k \in \mathcal{K}$. 
Accordingly,  the decoding order of
SIC at CUs can be expressed as
\begin{equation}
\small
    | \mathbf{h}_k \mathbf{w}_{K} |^2 \leq \cdots \leq | \mathbf{h}_k \mathbf{w}_{1} |^2 \leq | \mathbf{h}_k \mathbf{w}_{t,l} |^2\label{12}.
\end{equation}
According to Eqs. (\ref{9}) and (\ref{12}), the SINR of the sensing signal received at the CU $k$ can be expressed as
\begin{equation}
\small
    \text{SINR}_k^t = \frac{\left| \mathbf{h}_{k} \mathbf{w}_{t,l} \right|^2}{\sum_{m=1}^{K} \left| \mathbf{h}_{k} \mathbf{w}_{m} \right|^2 + \sigma_k^2}.
\end{equation}
Next, let $\gamma_t$ denote the predefined received SINR threshold for the sensing signal. According to the SIC decoding order at the CUs described in Eq. (\ref{12}), if $\text{SINR}_k^t \ge \gamma_t, \ k \in \mathcal{K}$, the interference from the sensing signal can be fully eliminated at all CUs by adopting the SIC technique in Eq. (\ref{9}). Accordingly, the SINR of the desired communication signal received at the $k$-th CU can be expressed as
\begin{equation}
\small
    \text{SINR}_k = \frac{\left| \mathbf{h}_{k} \mathbf{w}_{k} \right|^2}{\sum_{m=k+1}^{K} \left| \mathbf{h}_{k} \mathbf{w}_{m} \right|^2 + \sigma_k^2}.
\end{equation}

\subsection{Sensing Performance Metric}
We adopt the CRLB as the sensing performance metric. The unknown channel parameters to be estimated are contained in $\mathbf{v}=\begin{bmatrix}\theta_{ft} , \varphi_{ft}\end{bmatrix}^T$. Based on the stacked PRS observations over $L$ time slots $\mathbf{y}_t = [y_{t,1}, y_{t,2}, \dots, y_{t,L}]^T \in \mathbb{C}^{L \times 1}$  and the complex Gaussian distributed additive noise, we introduce the noise-free signal vector aggregated over $L$ time slots as:
\begin{align}
\small
\begin{split}
    &\boldsymbol{\mu}(\mathbf{v}) = \\&\Big[\sum_{k=1}^K \mathbf{g}_s\mathbf{w}_{k}s_{k,1}+\mathbf{g}_s\mathbf{w}_{t,1}\varphi_1,\ldots,\sum_{k=1}^K \mathbf{g}_s\mathbf{w}_{k}s_{k,L}+\mathbf{g}_s\mathbf{w}_{t,L}\varphi_L\Big]^{T}\\
    & = [\mathbf{g}_s\mathbf{W}_1\mathbf{s}_1,\ldots,\mathbf{g}_s \mathbf{W}_L\mathbf{s}_{L}]^T\\
    & = (\mathbf{I}_L\otimes\mathbf{g}_s)\cdot\mathbf{s}_{\text{stack}},
\end{split}
\end{align}
where $\mathbf{W}_l = [\mathbf{w}_{1},\ldots,\mathbf{w}_{K},\mathbf{w}_{t,l}]$, $\mathbf{I}_L\in \mathbb{R}^{L\times L}$ is an identity matrix, and $\mathbf{s}_{\text{stack}}=[(\mathbf{W}_1\mathbf{s}_1)^{T},\ldots,(\mathbf{W}_{L}\mathbf{s}_{L})^T]^T\in\mathbb{C}^{ML\times1}$. Herein, $\mathbf{s}_l = [s_{1,l},s_{2,l},\ldots,s_{K,l}, \varphi_{l}]^T$. 
Then, the Fisher information matrix for $\mathbf{v}$ is given by \cite{2021Intelligent}
\begin{equation}
\small
    \mathbf{F}_{i,j} \triangleq \frac{2}{\sigma^2_t} \Re\left\{ \frac{\partial \boldsymbol{\mu}^H}{\partial \nu_i} \frac{\partial \boldsymbol{\mu}}{\partial \nu_j} \right\}, \quad i,j = 1,2.
\end{equation}
In addition, 
the CRLB matrix is the inverse of the Fisher information matrix $\mathbf{F}$.
In order to obtain a closed-form expression of the CRLB for the AoD
estimation, the Fisher information matrix can be written as
\begin{equation}
\small
    \mathbf{F} = \begin{bmatrix}
\mathbf{F}_{\theta\theta} & \mathbf{F}_{\theta \varphi} \\
\mathbf{F}_{\theta \varphi}^H & \mathbf{F}_{\varphi \varphi}
\end{bmatrix},
\end{equation}
in which the expressions for the sub-matrices of $\mathbf{F}$ can be derived according to Eqs. (\ref{eq:037})-(\ref{eq:043}) as presented in Appendix A.
Accordingly, we derive the CRLB for AoD estimation, which corresponds to the first diagonal entry of  $\mathbf{F}^{-1}$ \cite{2024Cram}:
\begin{align}
\small
    \mathrm{CRLB}(\theta_{ft}) = \frac{\mathbf{F}_{\varphi \varphi}}{\mathbf{F}_{\theta\theta} \mathbf{F}_{\varphi \varphi} - |\mathbf{F}_{\theta \varphi}|^2}.
\end{align}


%



\section{Problem Formulation and Proposed Alternating Optimization Framework}
%

\subsection{Problem Formulation}
Specifically, the problem to be addressed is to maximize the minimum SINR by jointly optimizing the transmit beamforming vectors at the BS and the coefficient matrix of the FIM, while satisfying the predefined SINR requirements of each CU. The constraints include element-wise constraints on the FIM, design constraints on the sensing beamformers, the CRLB constraint on the AoD estimation of the SU, SIC decoding order constraints, the total transmit power constraint, and the phase shift constraints. Accordingly, the entire optimization problem can be formulated as:
\begin{subequations}\small\label{eqn-30}
  \begin{align}
    &\max_{\{\mathbf{w}_{k}\}_{k\in\mathcal{K}},\mathbf{w}_{t,l},\boldsymbol{\Theta},\mathbf{\Delta d}^{t}} \ \min_{k\in\mathcal{K}} \quad \mathrm{SINR}_k ,\label{eqn:objective} \\
    &\text{s.t.} \quad | \mathbf{\Delta d}^{t} | \le 0.5\lambda, \quad \forall n, \label{eqn:constraint1} \\
    &\quad\quad\ \mathrm{SINR}_k \ge \gamma_k, \quad k \in \mathcal{K}, \label{eqn:constraint2} \\
    &\quad\quad\ \mathrm{SINR}_k^t \ge \gamma_t, \quad k \in \mathcal{K}, \label{eqn:constraint3} \\
    &\quad\quad\ \Gamma_s \ge \mathrm{CRLB}(\theta_{ft}), \label{eqn:constraint4} \\
    &\quad\quad\ \sum_{k=1}^{K} \| \mathbf{w}_{k} \|^2 + \| \mathbf{w}_{t,l} \|^2 \leq P_{\max}, \label{eqn:constraint5} \\
   &\quad\quad\ \text{and} \,(\ref{12}), \label{eqn:constraint7}
  \end{align}
\end{subequations}
where $\gamma_k$ denotes the minimum SINR requirement of the $k$-th CU, and $\gamma_t$ represents the predefined received SINR threshold for the sensing signal. Constraint (\ref{eqn-30}d) ensures that the CUs are not impaired by the interference from the sensing signal. $\Gamma_s$ means the upper bound threshold of the CRLB for AoD estimation. $P_{max}$ limits the maximum transmit power at BS. 
It is evident from the  problem (\ref{eqn-30}) that the transmit beamforming vectors, surface
deformation, and reflection coefficient matrix of FIM are mutually coupled, making the above max-min fairness optimization difficult to be resolved.

\subsection{Alternating Optimization Algorithm}

Note that the formulated max-min fairness problem (\ref{eqn-30}) is non-convex due to the coupled optimization variables (beamforming vectors {${\mathbf{w}_{k}}$}, $\mathbf{w}_{t,l}$, FIM reflection  matrices $\boldsymbol{\Theta}$, and deformation vector $\mathbf{\Delta d}^{t}$) and non-convex constraints (\ref{eqn-30}c)-(\ref{eqn-30}e),  (\ref{eqn-30}g). 
To solve this problem efficiently, we propose an AO algorithm that decomposes the original non-convex problem into three convex subproblems, which are solved iteratively until reaching a convergence.
Within each iteration, we optimize one set of variables while fixing the other two, which guarantees that the objective value is non-decreasing and converges to a Karush-Kuhn-Tucker (KKT) stationary solution of the original problem.

\subsubsection{Beamforming Vector Optimization}
Fix FIM parameters $\mathbf{\Theta}$, and $\mathbf{\Delta d}^{t}$. The optimization problem reduces to designing the communication and sensing beamformers to maximize the minimum user SINR under power, QoS, and sensing SINR constraints. Using the max-min convexification (auxiliary variable $\tau$) and SOCP convexification for fractional SINR, the subproblem is written as:
\begin{subequations}
\small
  \begin{align}
    &\max_{\{\mathbf{w}_{k}\}_{k=1}^K, \mathbf{w}_{t,l}, \tau} \quad \tau \label{eq:26a}, \\
    &\text{s.t.} \quad \frac{|\mathbf{h}_k \mathbf{w}_{k}|^2}{\sum_{m=k+1}^{K} |\mathbf{h}_k \mathbf{w}_{m}|^2 + \sigma_k^2} \ge \tau, \quad \forall k, \label{eq:26b} \\
    &\quad\quad\ \frac{|\mathbf{h}_k \mathbf{w}_{t,l}|^2}{\sum_{m=1}^{K} |\mathbf{h}_k \mathbf{w}_{m}|^2 + \sigma_k^2} \ge \gamma_t, \quad \forall k, \label{eq:26c} \\
    &\quad\quad\ \sum_{k=1}^{K} \|\mathbf{\mathbf{w}}_{k}\|^2 + \|\mathbf{\mathbf{w}}_{t,l}\|^2 \le P_{\max}, \label{eq:26d} \\
    &\quad\quad\ \text{SIC decoding order constraints (\ref{12}).} \label{eq:26e}
  \end{align}\label{123}
\end{subequations}

The original communication SINR constraints take a fractional form and are non-convex. Since the denominator of each constraint is composed of inter-user interference power and noise power and is always strictly positive, the fractional inequalities can be transformed into quadratic inequalities by cross-multiplication, which enables convexification of the original constraints
\begin{equation}
\small
    \| \mathbf{h}_k \mathbf{w}_{k} \|^2 \ge \tau \left( \sum\nolimits_{m=k+1}^{K} \| \mathbf{h}_k \mathbf{w}_{m} \|^2 + \sigma_k^2 \right),\quad \forall k.\label{32}
\end{equation}
Similarly, the convex optimization formulation for the sensing SINR can be derived in the same manner
\begin{equation}
\small
    \| \mathbf{h}_k \mathbf{w}_{t,l} \|^2 \ge \gamma_t \left( \sum\nolimits_{m=1}^{K} \| \mathbf{h}_k \mathbf{w}_{m} \|^2 + \sigma_k^2 \right),\quad \forall k.\label{33}
\end{equation}
For the convexification of SIC constraints, we adopt the successive convex approximation (SCA) technique. Specifically, we fix the solution obtained in the previous iteration as a constant, perform the first-order Taylor expansion for the convex function, followed by converting the nonlinear inequality into an affine linear inequality, thus yielding convex constraints
\begin{equation}
\small
    \| \mathbf{h}_k \mathbf{w}_{k} \|^2 \approx \| \mathbf{h}_k \overline{\mathbf{w}}_{k} \|^2 + 2\left( \overline{\mathbf{w}}_{k}^H \mathbf{h}_k^H \mathbf{h}_k \right) \left( \mathbf{w}_{k} - \overline{\mathbf{w}}_{k} \right),\label{tu1}
\end{equation}
\begin{equation}
\small
    \| \mathbf{h}_k \mathbf{w}_{t,l} \|^2 \approx \| \mathbf{h}_k \overline{\mathbf{w}}_{t,l} \|^2 + 2\left( \overline{\mathbf{w}}_{t,l}^H \mathbf{h}_k^H \mathbf{h}_k \right) \left( \mathbf{w}_{t,l} - \overline{\mathbf{w}}_{t,l} \right).\label{tu2}
\end{equation}
Substituting Eqs. (\ref{tu1}) and (\ref{tu2}) into Eq. (\ref{12}), we obtain the corresponding expression:
\begin{equation}
\small
    \Re\left\{ \boldsymbol{g}_k^H ( \mathbf{w}_{k} - \overline{\mathbf{w}}_{k} ) \right\} \leq \Re\left\{ \boldsymbol{g}_t^H ( \mathbf{w}_{t,l} - \overline{\mathbf{w}}_{t,l} ) \right\},\label{36}
\end{equation}
where $\mathbf{g}_k = 2\mathbf{h}_k^H \mathbf{h}_k\overline{\mathbf{w}}_{k}$, $\mathbf{g}_t = 2\mathbf{h}_k^H \mathbf{h}_k\overline{\mathbf{w}}_{t,l}$. $\overline{\mathbf{w}}_{k}$ and $\overline{\mathbf{w}}_{t,l}$ denote the variables from the previous iteration, which are fixed as constants. By only taking the real part of the complex-valued equation, we obtain the final affine convex constraint. Subsequently, we convexify the CRLB constraints for the optimization 
\begin{equation}
\small
    \mathbf{F} = \begin{bmatrix}
\mathbf{F}_{\theta\theta} & \mathbf{F}_{\theta \varphi} \\
\mathbf{F}_{\theta \varphi}^H & \mathbf{F}_{\varphi\varphi}
\end{bmatrix} \succeq 0,
\end{equation}
where the FIM is always positive definite. By algebraic manipulation, the non-convex CRLB constraints can be transformed into linear matrix inequalities (LMIs), thereby achieving convexification
\begin{equation}
\small
    \Gamma_s \mathbf{F} - \frac{\sigma_t^2}{2} \mathbf{E}_{22} \succeq 0,\label{38}
\end{equation}
where $\mathbf{E}_{22}$ represents the second component of the identity vector, $\mathbf{E}_{22} = \begin{bmatrix}
0 & 0 \\
0 & 1
\end{bmatrix}$.

\begin{figure}    
\centering    
\includegraphics[width=0.5\textwidth]{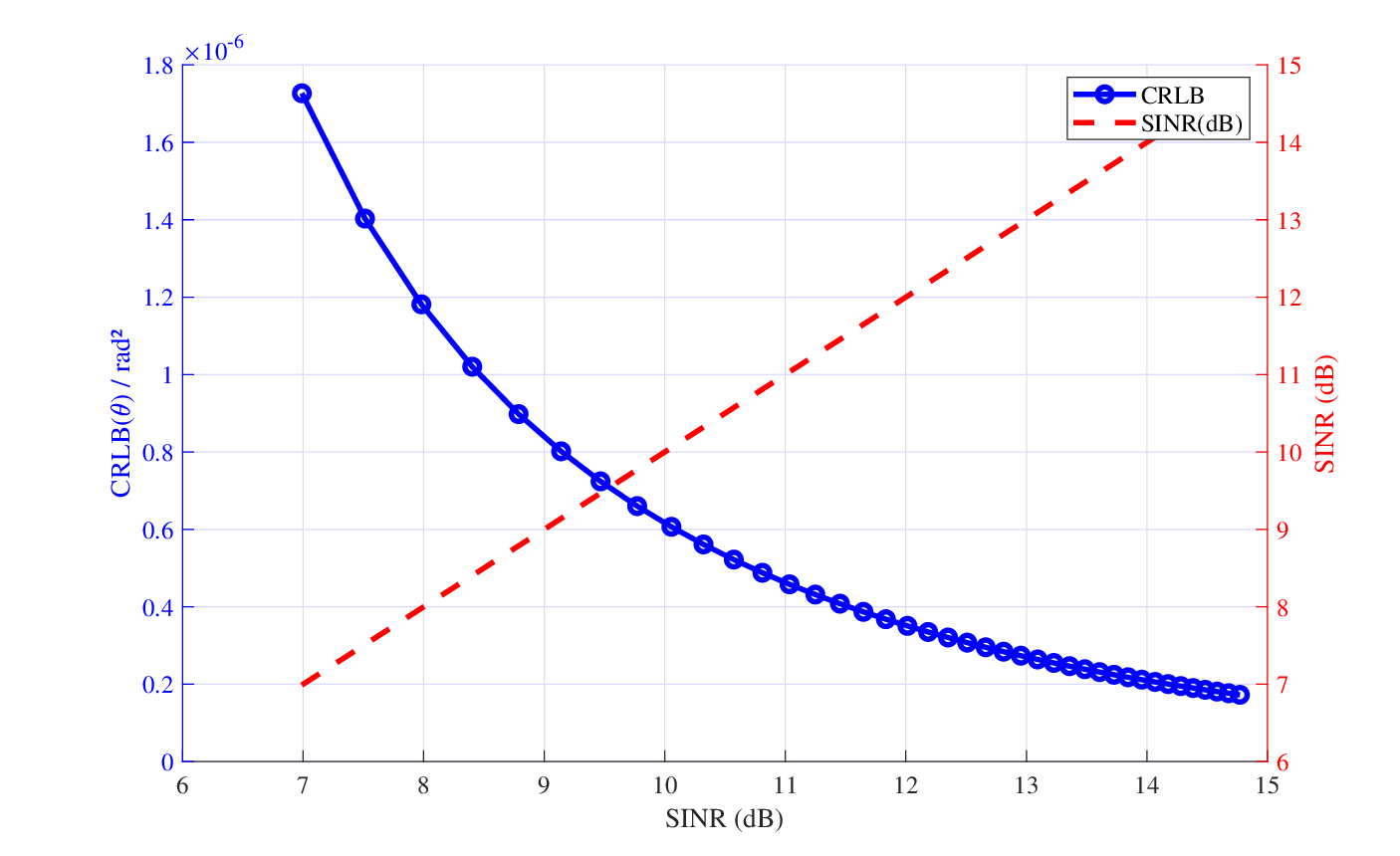}    
\caption{Communication SINR (dB) versus angular CRLB.}    
\label{fig:001}
\end{figure}

Consequently, the above optimization problem (\ref{123}) can be
approximately expressed as:\vspace{-0.3cm}
\begin{subequations}
\begin{align}   
\label{diancican}
&\max_{\{\mathbf{w}_k\}_{k=1}^K,\ \mathbf{w}_{t,l},\ \tau} \quad \tau \\ 
&\text{s.t.} \quad (\ref{eq:26d})-(\ref{eq:26e}),\ (\ref{32}),\ (\ref{33}),\ (\ref{36}),\ (\ref{38}),
\end{align}
\end{subequations}
which is convex and thus can be addressed by using some
existing tools, such as CVX\cite{2008CVX}.

\begin{figure*}[t]    
\centering
\begin{minipage}[t]{0.32\linewidth}        
\centering        
\includegraphics[width=1.1\linewidth,keepaspectratio]{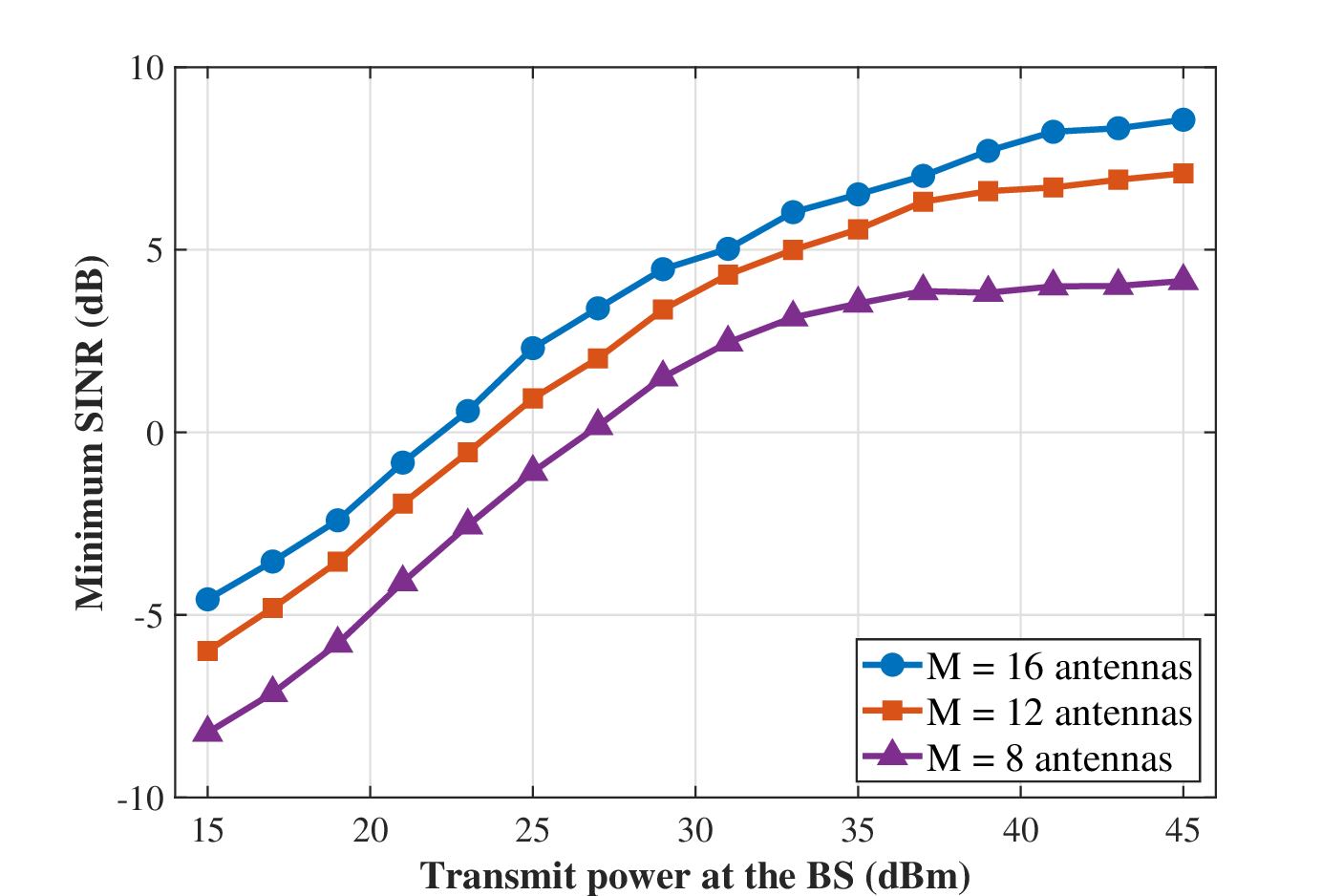} 
\caption{The ISAC fairness performance versus the transmit power at the BS employing the proposed approach with $N$ = 32 and varying antenna configuration $M$.}        
\label{fig:capacity_analysis}         
\end{minipage}    
\hfill    
\begin{minipage}[t]{0.32\linewidth}        
\centering        
\includegraphics[width=1.1\linewidth,keepaspectratio]{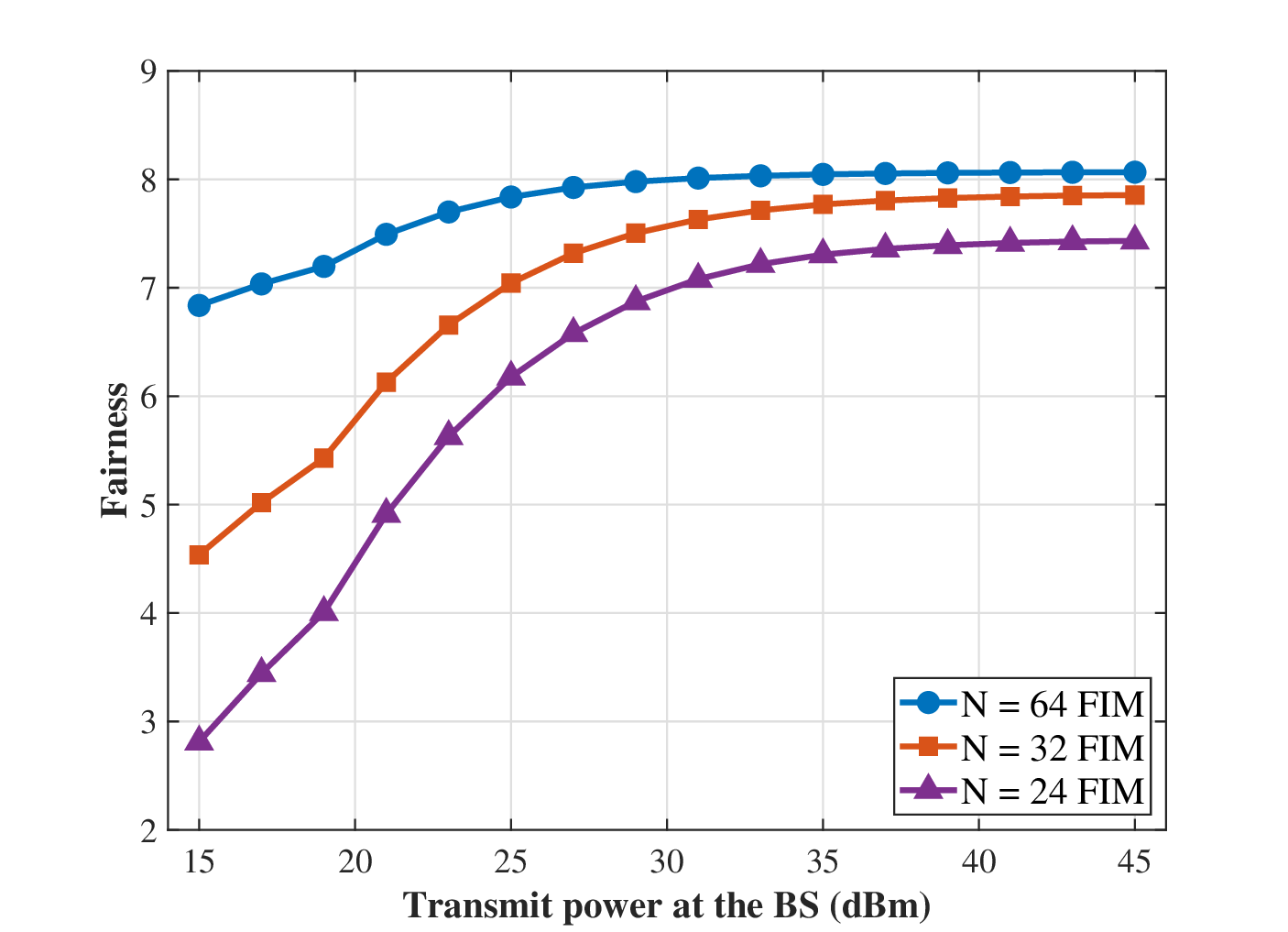}        
\caption{Fairness versus the transmit power using the proposed approach with $M$ = 16 and varying number of passive FIM elements.}         
\label{fig:impact_d20}          
\end{minipage}    
\hfill    
\begin{minipage}[t]{0.32\linewidth}        
\centering        
\includegraphics[width=1.1\linewidth,keepaspectratio]{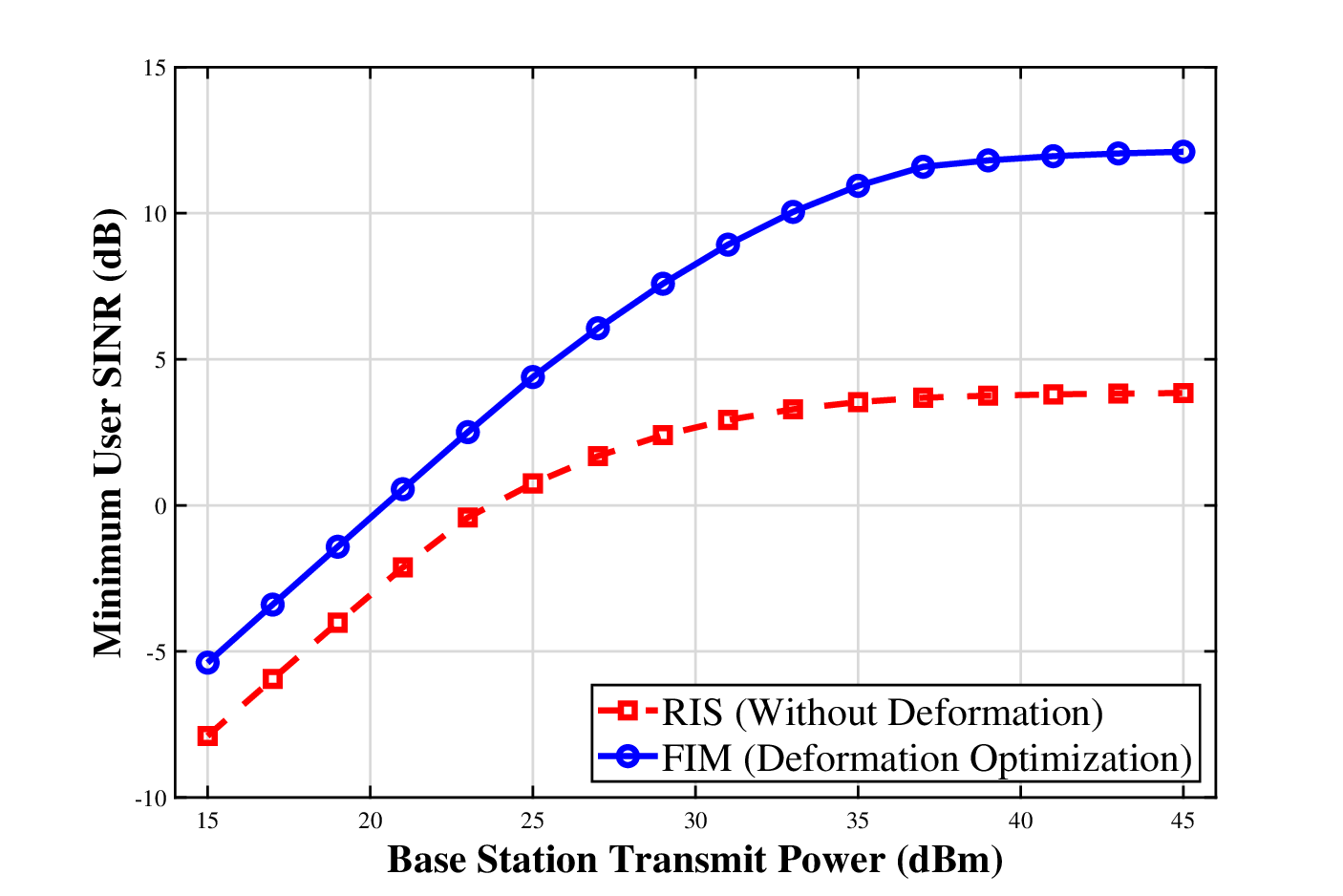}        
\caption{Achievable sum rate versus transmit power for FIM and RIS schemes under the system configuration of $N$ = 32 and $M$ = 16.}     
\label{fig:impact_d40_1}            
\end{minipage}
\end{figure*}

\subsubsection{Convex Optimization of FIM Electromagnetic Parameters}
We fix the beamforming vectors {$\mathbf{w}_{k}$}, $\mathbf{w}_{t,l}$ and the FIM deformation variable $\mathbf{\Delta d}^{t}$. In the derivation of the CRLB, we need to resolve the non-convex problem pertaining to electromagnetic parameter, we can derive the overall formulation of the present optimization problem. Therefore, this subsection addresses the non-convexity of FIM phase constraints and derives the closed-form solution for conjugate matching.

Substituting the diagonal FIM reflection matrix into the received signal model and expanding it into the element superposition form yield
$\mathbf{h}_{fk}^H(\mathbf{\Delta d}^t) \cdot \boldsymbol{\Theta} \cdot \mathbf{H}_{bf}^H(\mathbf{\Delta d}^t) \cdot \mathbf{w}_k=\sum_{n=1}^{N} g_n e^{j\phi_n}$.
Define the equivalent complex gain of the $n$-th FIM unit $\mathbf{g}_n = \mathbf{h}_{fk,n}(\mathbf{\Delta d}^t) \cdot \mathbf{H}_{bf,n}(\mathbf{\Delta d}^t) \cdot \mathbf{w}_k$. This gain is a constant in the current iteration. According to the complex triangle inequality, the modulus of the total signal achieves its global maximum when the phase factors of all units are conjugate matched with the equivalent complex gains.
Based on the above derivation and the theory presented in reference\cite{2019Intelligent}, 
the globally optimal closed-form solution for the FIM reflection phase can be achieved and can be expressed as
\begin{equation}
\small
    \left| \sum\nolimits_{n=1}^{N} \mathbf{g}_n e^{j\phi_{n}} \right| \leq \sum\nolimits_{n=1}^{N} |\mathbf{g}_n|,
\end{equation}
\begin{equation}
\small
    e^{j\phi_{n}^{\mathrm{opt}}} = \frac{\mathbf{g}_n^*}{|\mathbf{g}_n|},
\end{equation}
where $g_n^*$ is the complex conjugate of $g_n$. Take the argument of the optimal complex coefficient to obtain the closed-form solution of the phase shift of the $n$-th FIM unit:
\begin{equation}
\small
    \phi_{n}^{\mathrm{opt}} = \angle \left( \frac{g_n^*}{|g_n|} \right) = -\angle(g_n),
\end{equation}
\begin{equation}
\small
    \phi_{n}^{\mathrm{opt}} = -\angle\left( \mathbf{h}_{fk,\mathrm{n}}(\mathbf{\Delta d}^t) \cdot  \mathbf{H}_{bf,\mathrm{n}}(\mathbf{\Delta d}^t) \mathbf{w}_{k} \right).\label{29}
\end{equation}
Based on the above derivations, we update the optimization problem as follows:
\begin{subequations}  
\small\label{diancicanshu}
\begin{align}    
&\max_{\boldsymbol{\Theta}, \tau} \tau\label{eq:26a}, \\    
&\text{s.t.} (\ref{eq:26d})-(\ref{eq:26e}),(\ref{32}),\ (\ref{33}),\ (\ref{36}),\ (\ref{38}),(\ref{29}).
\end{align}
Obviously, the optimization problem (\ref{diancicanshu}) is convex and can be solved by CVX\cite{2008CVX}. With this approach, the optimal electromagnetic parameters of the FIM can be acquired with a low complexity.
\end{subequations}

\subsubsection{Optimizing the Normal Deformation Variable}
We fix the BS transmit beamforming vectors and FIM reflection coefficients in this subproblem. 
This surface deformation optimization is the most distinctive part of the proposed framework, which distinguishes FIM from traditional rigid RIS.

Different from conventional RIS with fixed planar geometry, each element of FIM can achieve independent displacement along the normal direction, introducing extra mechanical DoFs represented by $\mathbf{\Delta d}^{t}$. These tunable geometric parameters are directly embedded into the channel model and jointly optimized with communication and sensing constraints. The FIM deformation can dynamically reshape the array spatial structure, finely adjust the propagation phase of each wireless link, and further optimize channel alignment between the BS, FIM, CUs and the SU. Benefiting from the additional mechanical DoFs, the proposed deformation design brings performance gains that cannot be realized by phase optimization alone. 

In this subproblem, 
our proposed solution takes into account the following practical
hardware limitations: i) FIM deformation bounds, ii) multi-user communication fairness requirements, and iii) CRLB-based sensing accuracy constraints. All constraints have been transformed into convex forms, and the complete problem is formulated as:
\begin{subequations}  
\small
\begin{align}    
& \max_{\mathbf{\Delta d}^{t}, \tau} \tau\\    
&\text{s.t.} (\ref{eq:26d})-(\ref{eq:26e}),(\ref{32}),\ (\ref{33}),\ (\ref{36}),\ (\ref{38}).
\end{align}\label{youhuaxingbian}
\end{subequations}
Similarly, based on the above analysis, optimization problem (\ref{youhuaxingbian}) can also be solved via CVX \cite{2008CVX} to obtain the optimal deformation variables of the system.
\section{Performance Evaluations}
This section presents the simulation results of the proposed FIM-aided NOMA-ISAC system, which involves multiple CUs and a single SU. 
In our experiments, the number of CUs $K$ is set to be 3.
In addition, the default parameters are set as: orthogonal pilot sequences with fixed length $L=16$, carrier wavelength $\lambda=0.1\, m$, sensing noise power $\sigma^2_t=10^{-3}\, W$. 
Besides, the FIM reflection amplitudes satisfy the energy conservation constraint, while the conventional RIS (32 elements) uses fixed random phases and the proposed FIM adopts conjugate matching-based optimal phase and deformation. The receiver noise power is fixed at $10^{-5} \,W$, and we analyze the coupling between communication SINR and sensing CRLB, the impacts of BS antennas, FIM elements, and the fairness performance gain of FIM over traditional RIS.

Simulation results in Fig. \ref{fig:001} indicate that the CU's SINR has a significant negative correlation with the angular estimation CRLB, and exhibits a saturation characteristic with a fast-first and slow-later trend. At the low SINR range, improving communication performance can rapidly suppress noise and notably enhance sensing accuracy. When the SINR exceeds 10 dB, the sensing performance is constrained by inherent parameters such as array aperture and enters a saturation state. 

Fig. \ref{fig:capacity_analysis} plots the multi-user fairness performance in terms of minimum user SINR versus the BS transmit power under the FIM element number $N$=32 and different BS antenna configurations. Simulation results show that BS transmit power is positively correlated with multi-user fairness and exhibits a clear saturation trend. Fairness improves rapidly with power in 15–30 dBm, while performance saturates above 30 dBm due to inter-user interference and limited channel DoFs. 

Fig. \ref{fig:impact_d20} depicts the achievable sum-rate versus the BS transmit power under the BS antenna number $M$=16 and different numbers of FIM passive elements $N$. Simulation results demonstrate that increasing $N$ significantly improves multi-user fairness, especially in low-power regimes. As transmit power rises, the performance gap between different FIM element configurations shrinks and tends to saturate, since system performance is constrained by the intrinsic array gain at high power. 
Fig. \ref{fig:impact_d40_1} compares the multi-user fairness performance in terms of minimum user SINR versus the BS transmit power between the conventional rigid RIS and the proposed FIM scheme under the configuration of $N$=32 and $M$=16. It indicates that the multi-user fairness of both conventional RIS and the proposed FIM first rises rapidly and then gradually slows down with the increase of BS transmit power.

Thanks to the mechanical deformation of FIM, additional spatial DoFs are available for joint optimization. Compared with rigid RIS, FIM achieves better channel alignment and larger effective aperture, which suppresses multi-user interference and improves communication fairness. Meanwhile, the deformation design incorporated with geometry-aware CRLB optimization can guarantee high sensing accuracy for AoD estimation.

\section{Conclusion}
This paper has studied the user fairness optimization for FIM-aided NOMA-ISAC systems. We derived the closed-form CRLB for SU's parameters and integrated it into a max-min fairness framework to jointly optimize BS transmit beamforming, FIM reflection coefficients, and surface deformation. The non-convex optimization problem has been efficiently solved via an AO algorithm.
Simulation results confirmed the strong coupling between communication SINR and sensing CRLB, and demonstrated that our scheme significantly improves user fairness while balancing communication performance and sensing accuracy. 
This work provides a practical solution for fairness-aware FIM-ISAC system design, and provides useful insights for the deployment of 6G smart radio environments.

\begin{appendices}
\label{app:A}
\section{Closed-Form Expressions for Fisher Information Matrix and CRLB}
%
Note that the partial derivatives with respect to the elevation and azimuth AoD parameters $\theta_{ft}$ and  $\varphi_{ft}$ are expressed as
\begin{equation}
\scriptsize
\begin{split}
\label{eq:037}
    \frac{\partial \boldsymbol{\mu}}{\partial \theta_{ft}}  
    =& \Big(\mathbf{I}_{L} \otimes \Big(\big((\mathbf{D}_{11}\otimes\mathbf{I}_{N_x})\mathbf{h}_{ft}
    +(\mathbf{I}_{N_z} \otimes\mathbf{D}_{12})\mathbf{h}_{ft} \\&+\text{diag}(\boldsymbol{\alpha}_z \otimes \boldsymbol{\alpha}_x)\mathbf{D}_{21}\boldsymbol{\alpha}_{y}\big)^{H}\boldsymbol{\Theta}\mathbf{H}_{bf}\Big)\Big)\cdot \mathbf{s}_{\text{stack}},   
\end{split}
\end{equation}
\begin{equation}
\scriptsize
\begin{split}
\label{eq:038}
    \frac{\partial \boldsymbol{\mu}}{\partial \varphi_{ft}} = &\Big(\mathbf{I}_{L} \otimes \Big(\big(
    \text{diag}(\boldsymbol{\alpha}_y)(\mathbf{I}_{N_z} \otimes \mathbf{D}_{31})(\boldsymbol{\alpha}_z\otimes \boldsymbol{\alpha}_x) +
   \\ &\text{diag}(\boldsymbol{\alpha}_z \otimes \boldsymbol{\alpha}_x)\cdot \mathbf{D}_{32}\boldsymbol{\alpha}_{y}\big)^{H}\boldsymbol{\Theta}\mathbf{H}_{bf}\Big)\Big)\cdot \mathbf{s}_{\text{stack}}, 
\end{split}
\end{equation}
%
\begin{equation}
\scriptsize
\label{eq:039}
   \mathbf{D}_{11} = \mathrm{diag}\Big( jk_z\big(-\dot{N}_z \big)\sin\theta_{ft}, \dots, jk_z\big(\dot{N}_z\big)\sin\theta_{ft} \Big),
\end{equation}
%
\begin{equation}
\scriptsize
\label{eq:040}
   \mathbf{D}_{12} = \mathrm{diag}\Big( jk_x\big(-\dot{N}_x\big)\cos\theta_{ft}\cos\varphi_{ft}, \dots, jk_x\big( \dot{N}_x\big)\cos\theta_{ft}\cos\varphi_{ft} \Big),
\end{equation}
\begin{equation}
\scriptsize
\label{eq:041}
   \mathbf{D}_{21} = \mathrm{diag}\Big(-j\frac{2\pi \Delta d_{1}^{t}}{\lambda}\cos\theta_{ft}\sin\varphi_{ft}, \dots, -j\frac{2\pi \Delta d_{N}^{t}}{\lambda}\cos\theta_{ft}\sin\varphi_{ft} \Big),
\end{equation}
%
\begin{equation}
\scriptsize
\label{eq:042}
   \mathbf{D}_{31} = \mathrm{diag}\Big( -jk_x\big(-\dot{N}_x\big)\sin\theta_{ft}\sin\varphi_{ft}, \dots, -jk_x\big(\dot{N}_x\big)\sin\theta_{ft}\sin\varphi_{ft} \Big),
\end{equation}
\begin{equation}
\scriptsize
\label{eq:043}
   \mathbf{D}_{32} = \mathrm{diag}\Big( -j\frac{2\pi \Delta d_{1}^{t}}{\lambda}\sin\theta_{ft}\cos\varphi_{ft}, \dots, -j\frac{2\pi \Delta d_{N}^{t}}{\lambda}\sin\theta_{ft}\cos\varphi_{ft} \Big),
\end{equation}
where $k_z = \frac{2\pi d_z}{\lambda}$, $k_x=\frac{2\pi d_x}{\lambda}$, $\dot{N}_z = \frac{N_z-1}{2}$, and $\dot{N}_x = \frac{N_x-1}{2}$. 
Hence, based on the above Eqs. (\ref{eq:037})-(\ref{eq:043}),
the elements of Fisher information matrix can be derived as
$\mathbf{F}_{\theta\theta} = \frac{2}{\sigma^2_t} \Re\left\{ \frac{\partial \boldsymbol{\mu}^H}{\partial \theta_{ft}} \frac{\partial \boldsymbol{\mu}}{\partial \theta_{ft}} \right\}$, $\mathbf{F}_{\theta \varphi} =\frac{2}{\sigma^2_t} \Re\left\{ \frac{\partial \boldsymbol{\mu}^H}{\partial \theta_{ft}} \frac{\partial \boldsymbol{\mu}}{\partial \varphi_{ft}} \right\} $, and $\mathbf{F}_{\varphi \varphi}  = \frac{2}{\sigma^2_t} \Re\left\{ \frac{\partial \boldsymbol{\mu}^H}{\partial \varphi_{ft}} \frac{\partial \boldsymbol{\mu}}{\partial \varphi_{ft}} \right\}$.
Subsequently, we derive the CRLB for AoD estimation, which corresponds to the first diagonal entry of the inverse matrix of 
$\mathbf{F}$.

\end{appendices}

\small
\bibliographystyle{IEEEtran}%
\bibliography{bibfile}

@article{2016Next,
  title={Next Generation 5{G} Wireless Networks: A Comprehensive Survey},
  author={ Agiwal, Mamta  and  Roy, Abhishek  and  Saxena, Navrati },
  journal={IEEE
Commun. Surveys Tuts.},
  volume={18},
  number={3},
  pages={1617-1655},
  year={2016},
}

@article{2023Digital,
  title={Digital Twin-Empowered Communications: A New Frontier of Wireless Networks},
  author={ Bariah, Lina  and  Sari, H.  and  Debbah, M. },
  journal={ArXiv},
  volume={abs/2307.00973},
  year={2023},
}

@article{2020Smart,
  title={Smart Radio Environments Empowered by Reconfigurable Intelligent Surfaces: How it Works, State of Research, and Road Ahead},
  author={ Renzo, Marco Di  and  Zappone, Alessio  and  Debbah, Merouane  and  Alouini, Mohamed Slim  and  Tretyakov, Sergei },
  journal={IEEE J. Sel. Areas Commun.},
  volume={38},
  number={11},
  pages={2450-2525},
  year={2020},
}

@article{2021Cram,
  title={Cramér-{R}ao Bound Optimization for Joint Radar-Communication Beamforming},
  author={ Liu, Fan  and  Liu, Ya Feng  and  Li, Ang  and  Masouros, C.  and  Eldar, Yonina C. },
  journal={IEEE Trans. Signal Process.},
  volume={70},
  pages={240-253},
  year={2021},
}

@article{2019Intelligent,
  title={Intelligent Reflecting Surface Enhanced Wireless Network via Joint Active and Passive Beamforming},
  author={ Wu, Qingqing  and  Zhang, Rui },
  journal={IEEE Trans. Wireless Commun.},
  volume={18},
  number={11},
  pages={5394-5409},
  year={2019},
}

@article{2016Power,
  title={Power-Domain Non-Orthogonal Multiple Access ({NOMA}) in 5{G} Systems: Potentials and Challenges},
  author={ Islam, S. M. Riazul  and  others },
  journal={IEEE
Commun. Surveys Tuts.},
  volume={19},
  number={2},
  pages={721-742},
  year={2016},
}

@article{2017A,
  title={A Survey on Non-Orthogonal Multiple Access for 5{G} Networks: Research Challenges and Future Trends},
  author={ Ding, Zhiguo  and  others },
  journal={IEEE J. Sel. Areas Commun.},
  volume={35},
  number={10},
  pages={2181-2195},
  year={2017},
}

@article{2024Cram,
  title={Cramér-Rao Lower Bound and Fairness Optimization in {STAR}-{RIS} Assisted {ISAC} Systems},
  author={ Wang, Ruidong  and  Cao, Yuwen  and  Ohtsuki, Tomoaki  and  He, Jiguang },
  journal={2024 IEEE 100th Vehicular Technology Conference (VTC2024-Fall)},
  pages={1-6},
  year={2024},
}

@article{2021Intelligent,
  title={Intelligent Reflecting Surface Aided Dual-Function Radar and Communication System},
  author={ Elmeligy, Mohamed Rihan  and  Zhang, Peichang  and  Huang, Lei  and  Mohamed, Ehab Mahmoud },
  journal={IEEE Syst. J.},
  pages={475-486},
 volume={16},
  number={1},
  year={2021},
}

@article{2008CVX,
  title={CVX: MATLAB software for disciplined convex programming},
  author={ Grant, M. },
  journal={http://cvxr.com/cvx},
  year={2008},
}

@ARTICLE{11145277,
  author={Cao, Yuwen and Wu, Xiaowen and He, Jiguang and Ohtsuki, Tomoaki and Quek, Tony Q. S.},
  journal={IEEE Trans. Wireless Commun.}, 
  title={Coverage and Rate Performance Analysis of Multi-{RIS}-Assisted Dual-Hop mm{W}ave Networks}, 
  year={2026},
  volume={25},
  number={},
  pages={3137-3152},
  doi={10.1109/TWC.2025.3601815}}

@ARTICLE{10146432,
  author={Cao, Yuwen and Ohtsuki, Tomoaki and Maghsudi, Setareh and Quek, Tony Q. S.},
  journal={IEEE Trans. Veh.
Technol.}, 
  title={Deep Learning and Image Super-Resolution-Guided Beam and Power Allocation for mm{W}ave Networks}, 
  year={2023},
  volume={72},
  number={11},
  pages={15080-15085},
  doi={10.1109/TVT.2023.3282429}}

@ARTICLE{9314253,
  author={Echigo, Haruhi and Cao, Yuwen and Bouazizi, Mondher and Ohtsuki, Tomoaki},
  journal={IEEE Trans. Veh.
Technol.}, 
  title={A Deep Learning-Based Low Overhead Beam Selection in mm{W}ave Communications}, 
  year={2021},
  volume={70},
  number={1},
  pages={682-691},
  doi={10.1109/TVT.2021.3049380}}

@ARTICLE{11575614,
  author={Cao, Yuwen and Ohtsuki, Tomoaki and Maghsudi, Setareh and Quek, Tony Q. S.},
  journal={IEEE Trans. Cogn. Commun. Netw.}, 
  title={Memristor-Based Lightweight Meta Learning for Beam Prediction in Non-Stationary Environments}, 
  year={2026},
  volume={12},
  number={},
  pages={9401-9414},
  doi={10.1109/TCCN.2026.3706535}}

@ARTICLE{11060929,
  author={Teng, Zihao and others},
  journal={IEEE Trans. Veh.
Technol.}, 
  title={Flexible Intelligent Metasurface for Enhancing Multi-Target Wireless Sensing}, 
  year={2025},
  volume={74},
  number={12},
  pages={19825-19830},
  doi={10.1109/TVT.2025.3584865}}
\end{document}